\documentclass{sig-alternate-05-2015}
\renewcommand{\vec}[1]{\mathbf{#1}}
\usepackage{amsmath}
\usepackage{threeparttable}
\usepackage{multirow} 
\usepackage{graphicx}
\usepackage{subfigure}
\usepackage[justification=centering]{caption}

\begin{document}
%
%\CopyrightYear{2016}
%\setcopyright{acmlicensed}
%\conferenceinfo{CIKM'16 ,}{October 24 - 28, 2016, Indianapolis, IN, USA}
%\isbn{978-1-4503-4073-1/16/10}\acmPrice{\$15.00}
%\doi{http://dx.doi.org/10.1145/2983323.2983649}

% DOI
%\doi{10.475/123_4}
%
%% ISBN
%\isbn{123-4567-24-567/08/06}
%
%%Conference
%\conferenceinfo{CIKM '13}{June 16--19, 2013, Seattle, WA, USA}
%
%\acmPrice{\$15.00}

%
% --- Author Metadata here ---
%\conferenceinfo{CIKM}{2016 Indianapolis, USA}
%\CopyrightYear{2007} % Allows default copyright year (20XX) to be over-ridden - IF NEED BE.
%\crdata{0-12345-67-8/90/01}  % Allows default copyright data (0-89791-88-6/97/05) to be over-ridden - IF NEED BE.
% --- End of Author Metadata ---

\title{Top-N Recommendation on Graphs}
%\subtitle{[Extended Abstract]
%\titlenote{A full version of this paper is available as
%\textit{Author's Guide to Preparing ACM SIG Proceedings Using
%\LaTeX$2_\epsilon$\ and BibTeX} at

\numberofauthors{1} %  in this sample file, there are a *total*

\author{
% 1st. author
\alignauthor
Zhao Kang, Chong Peng, Ming Yang, Qiang Cheng\\
       \affaddr{Department of Computer Science, Southern Illinois University, Carbondale, IL, USA}\\
%       \affaddr{1932 Wallamaloo Lane}\\
%       \affaddr{Wallamaloo, New Zealand}\\
       \email{\{Zhao.Kang, pchong, ming.yang, qcheng\}@siu.edu}
%% 2nd. author
%\alignauthor
%G.K.M. Tobin\titlenote{The secretary disavows
%any knowledge of this author's actions.}\\
%       \affaddr{Institute for Clarity in Documentation}\\
%       \affaddr{P.O. Box 1212}\\
%       \affaddr{Dublin, Ohio 43017-6221}\\
%       \email{webmaster@marysville-ohio.com}
%% 3rd. author
%\alignauthor Lars Th{\o}rv{\"a}ld\titlenote{This author is the
%one who did all the really hard work.}\\
%       \affaddr{The Th{\o}rv{\"a}ld Group}\\
%       \affaddr{1 Th{\o}rv{\"a}ld Circle}\\
%       \affaddr{Hekla, Iceland}\\
%       \email{larst@affiliation.org}
%\and  % use '\and' if you need 'another row' of author names
%% 4th. author
%\alignauthor Lawrence P. Leipuner\\
%       \affaddr{Brookhaven Laboratories}\\
%       \affaddr{Brookhaven National Lab}\\
%       \affaddr{P.O. Box 5000}\\
%       \email{lleipuner@researchlabs.org}
% 5th. author
%\alignauthor Sean Fogarty\\
%       \affaddr{NASA Ames Research Center}\\
%       \affaddr{Moffett Field}\\
%       \affaddr{California 94035}\\
%       \email{fogartys@amesres.org}
%% 6th. author
%\alignauthor Charles Palmer\\
%       \affaddr{Palmer Research Laboratories}\\
%       \affaddr{8600 Datapoint Drive}\\
%       \affaddr{San Antonio, Texas 78229}\\
%       \email{cpalmer@prl.com}
}
% There's nothing stopping you putting the seventh, eighth, etc.
% author on the opening page (as the 'third row') but we ask,
% for aesthetic reasons that you place these 'additional authors'
% in the \additional authors block, viz.
%\additionalauthors{Additional authors: John Smith (The Th{\o}rv{\"a}ld Group,
%email: {\texttt{jsmith@affiliation.org}}) and Julius P.~Kumquat
%(The Kumquat Consortium, email: {\texttt{jpkumquat@consortium.net}}).}
% Just remember to make sure that the TOTAL number of authors
% is the number that will appear on the first page PLUS the
% number that will appear in the \additionalauthors section.

\maketitle
\begin{abstract}
Recommender systems play an increasingly important role in online
applications to help users find what they need or prefer. Collaborative filtering algorithms that generate predictions by analyzing the user-item rating matrix perform poorly when the matrix is sparse. To alleviate this problem, 
this paper proposes a simple recommendation algorithm that fully exploits the similarity information among users and items and intrinsic structural information of the user-item matrix. The proposed method constructs a new representation which preserves affinity and structure information in the user-item rating matrix and then performs recommendation task.  
To capture proximity information about users and items, two graphs are constructed. Manifold learning idea is used to constrain the new representation to be smooth on these graphs, so as to enforce users and item proximities. Our model is formulated as a convex optimization problem, for which we need to solve the well known Sylvester equation only. We carry out extensive empirical evaluations on six benchmark datasets to show the effectiveness of this approach.%, and observe that the proposed approach significantly outperforms the state-of-the-art methods. This indicates the great potential of this simple yet highly effective method in real-life application.
\end{abstract}

%\ccsdesc[500]{Information systems~Recommender systems}
%\ccsdesc[500]{Information systems~Personalization}
%%\ccsdesc[500]{Human-centered computing~Collaborative filtering}
%\ccsdesc[300]{Computing methodologies~Regularization}


%
%  Use this command to print the description
%
\printccsdesc

% We no longer use \terms command


\section{Introduction}
%Recommender systems \cite{adomavicius2005toward} have become increasingly indispensable in many applications, such as movie, music, news, book, and video recommendations. They attempt to identify and recommend items that are likely to arouse users' interests. Specifically, the task of Top-$N$ recommendation is to produce size-$N$ ranked lists of items that best fit customers' personal tastes and special needs. Existing recommender systems can be roughly divided into two categories: content-based \cite{balabanovic1997fab} and collaborative filtering (CF) based \cite{breese1998empirical} methods. 

%Content-based methods make use of profiles of users or items to characterize their nature. More precisely, they identify the common features of items that have received favorable ratings from a user, and then recommend this user new items that share these features. This approach suffers from the problems of limited content analysis and over-specialization \cite{shardanand1995social}. Limited content analysis occurs when there is not enough information about the users or the items. For instance, it is difficult to parse semantic contents of musics and videos. On the other hand, over-specialization is an inherent side effect. An item that is similar to the ones liked by the user would be highly recommended. As a result, this method usually fails to recommend items that are different but may still arouse the user's interest. 

Recommender systems have become increasingly indispensable in many applications \cite{adomavicius2005toward}. Collaborative filtering (CF) based methods are a fundamental building block in many recommender systems. CF based recommender systems predict the ratings of items to be given by a user based on the ratings of the items previously rated by other users who are most similar to the target user. %Compared to content-based methods, CF based methods act only on a user-item rating matrix which represents user explicit (e.g., rating/review) or implicit (e.g. click-through, transaction, and check-in) feedback information. Besides, CF based methods have the capability of exposing unexpected items to users. Therefore, CF based methods are more popular than content-based ones and  Many hybrid methods, which integrate the above two categories of algorithms, have also been developed \cite{basilico2004unifying}.

CF based methods can be classified into memory-based methods
\cite{sarwar2001item} and model-based methods \cite{kang2016top,koenigstein2012efficient}. The former 
includes two popular methods, user-oriented \cite{herlocker1999algorithmic} and item-oriented, e.g., ItemKNN  \cite{deshpande2004item}, depending on whether the neighborhood information is derived from similar users or items. First they compute similarities between the active item and other items, or between the active user and other users. Then they predict the unknown rating by combining the known rating of top $k$ neighbors. %Work \cite{wang2006unifying} combines them and the final rating is predicted based on three sources: predictions based on ratings of the same item by other users, predictions based on different item ratings made by the same user, and predictions based on data from similar users' ratings of other but similar items. 
Due to the simplicity of memory-based CF, it has been successfully applied in industry. However, it suffers from several problems, including data sparsity, cold start and data correlation \cite{cacheda2011comparison}, as users typically rate only a small portion of the available items, and they also tend to rate similar items closely. Therefore, the similarities between users or items cannot be accurately obtained with the existing similarity measures such as cosine and Pearson correlation, which in turn compromises the recommendation accuracy.  

To alleviate the problems of memory-based methods, many model-based methods have been proposed, which use observed ratings to learn a predictive model. %Methods in this category are numerous, including latent semantic analysis \cite{hofmann2003collaborative}, Bayesian clustering \cite{breese1998empirical}, latent dirichlet allocation \cite{blei2003latent}, and matrix factorization (MF) based methods \cite{koren2008factorization}.  %srebro2003weighted,mnih2007probabilistic,bell2007modeling,koren2008factorization,paterek2007improving, takacs2009scalable}. 
Among them, matrix factorization (MF) based models, e.g., PureSVD \cite{cremonesi2010performance} and weighted regularized MF (WRMF) \cite{hu2008collaborative}, are very popular due to their capability of capturing the implicit relationships among items and their outstanding performance. Nevertheless, it introduces high computational complexity and also faces the problem of uninterpretable recommendations. Because the rating matrix is sparse, the factorization of the user-item matrix may lead to inferior solutions \cite{gu2010collaborative}. By learning an aggregation coefficient matrix \cite{kang2015robust}, recently, sparse linear method (SLIM) \cite{ning2011slim} has been proposed and shown to be effective. However, it just captures relations between items that have been co-purchased/co-rated by at least one user \cite{kang2016sdm}. Moreover, it only explores the linear relations between items. Another class of methods use Bayesian personalized ranking (BPR) criterion to measure the difference between the rankings of user-purchased items and the remaining items. For instance, BPRMF and BPRKNN \cite{rendle2009bpr} have been demonstrated to be effective for implicit feedback datasets.

In this paper, we propose a novel Top-$N$ recommendation model based on graphs. This method not only takes into account the neighborhood information, which is encoded by our user graph and item graph, but also reveals hidden structure in the data by deploying graph regularization. In the real world, data often reside on low-dimensional manifolds embedded in a high-dimensional ambient space. Like the Netflix Prize problem, where the size of the user-item matrix can be huge, there exist relationships between users (such as their age, hobbies, education, etc.) and movies (such as their genre, release year, actors, origin country, etc.). Moreover, people sharing the same tastes for a class of movies are likely to rate them similarly. As a result, the rows and columns of the user-item matrix possess important structural information, which should be taken advantage of in actual applications. 

To preserve local geometric and discriminating structures embedded in a high-dimensional space, numerous manifold learning methods, such as locally linear embedding (LLE) \cite{roweis2000nonlinear}, locality preserving projection (LPP) \cite{niyogi2004locality}, have been proposed. In recent years, graph regularization based non-negative matrix factorization \cite{cai2011graph} of data representation has been developed to remedy the failure in representing geometric structures in data. Inspired by this observation, to comprehensively consider the associations between the users and items and the local manifold structures of the user-item data space, we propose to apply both user and item graph regularizations. Unlike many existing recommendation algorithms, we first establish a new representation which is infused with the above information. It turns out that this new representation is not sparse anymore. Therefore, we perform recommendation task with this novel representation. %Empirical analysis on various real datasets demonstrates that this approach can improve recommendation quality considerably. 

%In summary, our main contributions in this paper lie in the following three aspects:
%\begin{enumerate}
%\item We propose a novel recommendation algorithm by constructing a new representation that preserves proximities between users and items and manifold structures of the user-item matrix. Fully exploring those information will not only improves the recommendation quality but also alleviates  sparsity issue. More specifically, as the data manifolds are unknown a priori, the local geometric structures of data are modeled by two nearest neighbour graphs and then the graph structures are utilized to find the new representation of the user-item matrix.
%\item Our model is convex and can be solved easily by the well known Sylvester equation. Compared to popular MF-based methods, it is guaranteed to have the global optimal solution and enjoys the benefit of efficiency. %These make it have great potential for real-time applications.
%\item Experiments on six benchmark datasets demonstrate that our proposed method outperforms other state-of-the-art algorithms.%, which give similar performances on most datasets. 
%\end{enumerate}
%The remainder of this paper is organized as follows. In Section \ref{notation}, we define some notations. Section \ref{proposed} introduces the proposed method and its solution. Section \ref{related} describes the connection of our method to existing work. In Section \ref{expeval}, we describe our experimental framework. Experimental results and analysis are presented in Section \ref{discuss}. Section \ref{conclude} draws conclusions. 

\section{Definitions and notations}
\label{notation}
Let $U=\{u_1,u_2,...,u_m\}$ and $T=\{t_1,t_2,...,t_n\}$ represent the sets of all users and all items, respectively. The whole set of user-item purchases/ratings are represented by the user-item matrix $X$ of size $m\times n$. Element $x_{ij}$ is 1 or a positive value if user $u_i$ has ever purchased/rated item $t_j$, otherwise it is marked as $0$. The $i$-th row of $X$ denotes the purchase/rating history of user $u_i$ on all items. The $j$-th column of $X$ is the purchase/rating history of all users on item $t_j$. $\|X\|_F^2=\sum_{i}\sum_j x_{ij}^2$ is the squared Frobenius norm of $X$. Tr$(\cdot)$ stands for the trace operator. $I$ denotes the identity matrix. $\odot$ is the Hadamard product. %The aggregation coefficient matrix is represented by $W$. $\vec{w}_j$ is a sparse size-$n$ column vector of aggregation coefficients. $\|W\|_1=\sum\limits_{i}\sum\limits_j|w_{ij}|$ is the entry-wise $l_1$-norm of $W$. $\|W\|_F^2=\sum\limits_{i}\sum\limits_j w_{ij}^2$ is the squared Frobenius norm of $W$. The nuclear norm of $W$ is $\|W\|_*=\sum_{i=1}^{rank(W)}\sigma_i$, where $\sigma_i$ is the $i$-th singular value of $W$. Tr$(\cdot)$ stands for the trace operator. $I$ denotes the identity matrix.
%
%In this paper, all vectors (e.g., $\vec{x}_i$, $\vec{x}_j$) are denoted by bold lower case letters. All matrices (e.g. $W$, $X$) are represented by upper case letters. A predicted value is denoted by having a $\wedge$ mark.
\section{Proposed Model}
\label{proposed}
%In this section, we propose a novel Top-$N$ recommendation method by fully exploiting the structure information hidden in the rows and columns of the user-item matrix. We reconstruct a new representation $Y\in\mathcal{R}^{m\times n}$ from the user-item rating matrix $X$ while simultaneously preserving affinity and structure information of $X$.
\subsection{User and Item Graphs}
User-item rating matrix is an overfit representation of user tastes and item descriptions. This leads to problems of synonymy, computational complexity, and potentially poorer results. Therefore, a more compact representation of user tastes and item descriptions is preferred. 
Graph regularization is effective in preserving local geometric and discriminating structures embedded in a high-dimensional space. %has been widely used including dimensionality reduction \cite{yan2007graph} and clustering \cite{cai2008non}. 
It is based on the well known manifold assumption \cite{niyogi2004locality}: If two data points such as $\vec{x}_i$ and $\vec{x}_j$ are close in the geodesic distance on the data manifold, then their corresponding representations $\vec{y}_i$ and $\vec{y}_j$ are also close to each other. In practice, it is difficult to accurately estimate the global manifold structure of the data due to the insufficient number of samples and the high dimensionality of the ambient space. Therefore, many methods resort to local manifold structures. Much effort on manifold learning \cite{niyogi2004locality} has shown that local geometric structures of the data manifold can be effectively modeled through a nearest neighbor graph on sampled data points. 

We adopt graph regularization to incorporate user and item proximities. In this paper, we construct two graphs: the user graph and the item graph. We assume that users having similar tastes for items form communities in the user graph, while items having  similar appeals to users form communities in the item graph. Since ``birds of a feather flock together", this assumption is plausible and turns out to indeed benefit recommender systems substantially in our experiment results. As an example for movie recommendation, the users are the vertices of a ``social graph" whose edges represent relations induced by similar tastes. 

More formally, we construct an undirected weighted graph $G_c=(V_c,E_c;S_c)$ on items, called the item graph. The vertex set $V_c$ corresponds to items $\{t_1,\cdots,t_n\}$ with each node $t_i$ corresponding to a data point $\vec{x}_i$ which is the $i$-th column of $X$. Symmetric adjacency matrix $S_c$ encodes the inter-item information, in which $s_{ij}$ is the weight of the edge joining vertices $t_i$ and $t_j$ and represents how strong the relationship or similarity items $t_i$ and $t_j$ have. $E_c=\{e_{ij}\}$ is the edge set with each edge $e_{ij}$ between nodes $t_i$ and $t_j$ associated with a weight $s_{ij}$. The graph regularization on the item graph is formulated as 
\scriptsize
\begin{equation}
\label{manifold}
\begin{split}
&\frac{1}{2}\sum_{i,j=1}^{n}\|\vec{y_i}-\vec{y}_j\|_2^2 s_{ij}=\sum_{i=1}^n\vec{y}_i^T\vec{y}_id_{ii}-\sum_{i,j=1}^{n}\vec{y}_i^T\vec{y}_j s_{ij}\\
=&\textrm{Tr}(YD_cY^T)-\textrm{Tr}(YS_cY^T)=\textrm{Tr}(YL_cY^T),
\end{split}
\end{equation}
\normalsize
where $D_c$ is a diagonal matrix with $d_{ii}=\sum_{j=1}^n s_{ij}$, and $L_c=D_c-S_c$ is the graph Laplacian. To preserve the structural information of the manifold, we want (\ref{manifold}) to be as small as possible. It is apparent that minimizing (\ref{manifold}) imposes the smoothness of the representation coefficients; i.e., if items $t_i$ and $t_j$ are similar (with a relatively bigger $s_{ij}$), their low-dimensional representations $\vec{y}_i$ and $\vec{y}_j$ are also close to each other. Therefore, optimizing (\ref{manifold}) is an attempt to ensure the manifold assumption. 
% \cite{chung1997spectral}

The crucial part of graph regularization is the definition of the adjacency matrix $S_c$. There exist a number of different similarity metrics in the literature \cite{sarwar2001item}, e.g., cosine similarity, Pearson correlation coefficient, and adjusted cosine similarity. For simplicity, in our experiment, we use cosine similarity for explicit rating datasets and Jaccard coefficient for implicit feedback datesets. For binary variables, the Jaccard coefficient is a more appropriate similarity metric than cosine because it is insensitive to the amplitudes of ratings. It measures the fraction of users who have interactions with both items over the number of users who have interacted either of them. 
Formally, according to cosine definition, the similarity $s_{ij}$ between two items $t_i$ and $t_j$ is defined as
%\begin{equation*}
$s_{ij}=\frac{\vec{x}_i\cdot\vec{x}_j}{\|\vec{x}_i\|_2\|\vec{x}_j\|_2}$, %s_{ij}=\frac{\vec{x}_i\cap\vec{x}_j}{\vec{x}_i\cup \vec{x}_j},
%\end{equation*}
where `$\cdot$' denotes the vector dot-product operation. For Jaccard coefficient, 
%\begin{equation*}
$s_{ij}=\frac{|\vec{x}_i\cap\vec{x}_j|}{|\vec{x}_i\cup \vec{x}_j|}$, 
%\end{equation*}
where $\cap$ and $\cup$ represent intersection and union operations, respectively. Likewise, by defining the user graph $G_r=(V_r,E_r;S_r)$ whose vertex set $V_r$ corresponds to users $\{u_1,\cdots,u_m\}$, we get a corresponding expression $\textrm{Tr}(Y^TL_rY)$. Here $L_r$ denotes the Laplacian of $G_r$, which is similarly obtained from the data points corresponding to the users, that is, the rows of $X$.  

\subsection{Model}
 By exploiting both user and item graphs, our proposed model can be written as
\scriptsize
\begin{equation}
\label{model}
\min_Y \|X-Y\|_F^2+\alpha \textrm{Tr}(YL_cY^T)+\beta \textrm{Tr}(Y^TL_rY).
\end{equation}
\normalsize
The first term of (\ref{model}) penalizes large deviations of the predictions from the given ratings. The last two terms measure the smoothness of the predicted ratings on the graph structures and encourage the ratings of nodes with affinity to be similar. They can alleviate the data sparsity issue to some extent. When the item neighborhood information is not available, user neighborhood information might exist, vice versa. The parameters $\alpha$ and $\beta$ adjust the balance between the reconstruction error and graph regularizations. 
%Communities of people share preferences, while products form clusters that receive similar ratings 

By setting the derivative of the objective function of (\ref{model}) with respect to $Y$ to zero, we have
%\begin{equation}
%Y-X+\alpha YL_c+\beta L_rY=0,
%\end{equation}
%or equivalently,
\scriptsize
\begin{equation}
\label{sylvester}
(\beta L_r+I)Y+\alpha YL_c=X.
\end{equation}
\normalsize
Equation (\ref{sylvester}) is the well known Sylvester equation, which costs $O(m^3)$ or $O(n^3)$ with a general solver. But in our situation, $X$ is usually extremely sparse, and $L_r$ and $L_c$ can also be sparse, especially for large $m$ and $n$, so the cost can be $O(m^2)$ or $O(n^2)$, or sometimes even as low as $O(m)$ or $O(n)$ \cite{benner2009adi}. Many packages or programs are available to solve (\ref{sylvester}). %We use the Matlab built-in function lyap or sylvester to solve it in this work.

To use the reconstructed matrix $Y$
to make recommendations
for user $u_i$, we just sort $u_i$'s non-purchased/non-rated items based
on their scores in non-increasing order and recommend the top $N$ items.

\section{Connection to existing work}
\label{related}
To the best of our knowledge, there are very few studies on graph Laplacian in the context of recommendation task. Graph regularized weighted nonnegative matrix factorization (GWNMF) \cite{gu2010collaborative} was proposed to incorporate the neighborhood information in a MF approach. It solves the following problem
\scriptsize
\begin{equation}
\begin{split}
&\min_{U,V} \|M\odot (X-UV^T)\|_F^2+\alpha Tr(U^TL_rU)+\beta Tr(V^TL_cV)\\
 &s.t.\quad U\ge 0, V\ge 0,
 \end{split}
\label{GWNMF}
\end{equation}
\normalsize
where $M$ is an indicator matrix. $U$ and $V^T$ are in latent spaces, whose dimensionality is usually specified with an additional parameter. The latent factors are generally not obvious and might not necessarily be interpretable or intuitively understandable. Here (\ref{GWNMF}) has to learn both user and item representations in the latent spaces. In our approach, we just need to learn one representation and thus the learning process is simplified. On the other hand, $U$ and $V^T$ are supposed to be of low dimensionality, and thus useful information can be lost during the low-rank approximation of $X$ from $U$ and $V^T$. The encoding of graph Laplacian on $U$ and $V^T$ might be not accurate any more. On the contrary, our method can better preserve the information in $X$, so it can potentially give better recommendations than GWNMF. Moreover, it is well known that several drawbacks exist in the MF approach, e.g., low convergence rate, many local optimums of $U$ and $V^T$ due to the non-convexity of (\ref{GWNMF}). In contrast, our model (\ref{model}) is strongly convex, admitting a unique, globally optimal solution. 
\begin{table}[ht]
\caption{The datasets used in evaluation}
\label{datainfo}
\renewcommand{\arraystretch}{.6}
\begin{center}
%\small
%\begin{sc}
\resizebox{.45\textwidth}{!}{
\begin{tabular}{llllllll}
\multicolumn{1}{c}{dataset}  &\multicolumn{1}{c}{\#users} &\multicolumn{1}{c}{ \#items}  &\multicolumn{1}{c}{\#trns} &\multicolumn{1}{c}{rsize}  &\multicolumn{1}{c}{csize}&\multicolumn{1}{c}{ density}&\multicolumn{1}{c}{ratings}\\
\hline\hline \\
Delicious&1300&4516&17550&13.50&3.89&0.29\%&-\\
 lastfm&8813&6038&332486&37.7&55.07&0.62\%&- \\
BX&4186&7733&182057&43.49&23.54&0.56\%&- \\
\hline \\
Filmtrust &1508&2071&35497&23.54&17.14&1.14\%&0.5-4\\
Netflix&6769&7026&116537&17.21&16.59&0.24\%&1-5 \\
Yahoo&7635&5252&212772&27.87&40.51&0.53\% &1-5 \\
\hline
\end{tabular}}
\begin{tablenotes}
     \tiny
      \item In this table, the ``\#users", ``\#items", ``\#trns" columns represent the number of users, number of items and number of transactions, respectively, in each dataset. The ``rsize" and ``csize" columns show the average number of ratings of each user and of each item, respectively, in each dataset. Column corresponding to ``density" shows the density of each dataset (i.e., density=\#trns/(\#users$\times$\#items)). The ``ratings" column is the rating range of each dataset . The ratings in FilmTrust are real values with step 0.5, while in the other datasets are integers.
    \end{tablenotes}
%\end{sc}
%\end{small}
\end{center}
\end{table}

\section{Experimental Evaluation}
\label{expeval}
% In this section, we discuss the evaluation of our algorithm, focusing on datasets, evaluation metrics, and the state-of-the-art algorithms. 
\subsection{Datasets}
%Many recommendation algorithms are designed for explicit feedback such as ratings. However, implicit feedback, in which a user's preferences are expressed through user history such as views or purchases, is arguably the most typical setting in real-world applications. It is much less explored than the explicit feedback. 

%To show the effectiveness of our proposed method in more general situations, we evaluate it on both explicit and implicit feedback datasets. The characteristics of the datasets are listed in Table \ref{datainfo}.
%
%As shown in Table \ref{datainfo}, Delicious, lastfm and BX have only implicit feedback (e.g., listening history); i.e., they are represented by binary matrices. In particular, Delicious was from the bookmarking and tagging information of 2$K$ users in Delicious social bookmarking system\footnote{http://www.delicious.com}, in which each URL was bookmarked by at least 3 users. Lastfm represents music artist listening information obtained from the last.fm online music system\footnote{ http://www.last.fm }, in which each music artist was listened to by at least 10 users and each user listened to at least 5 artists. BX is derived from the Book-Crossing dataset\footnote{http://www.informatik.uni-freiburg.de/~cziegler/BX/} such that only implicit interactions were contained and each book was read by at least 10 users. 
Table \ref{datainfo} shows the characteristics of the datasets. Delicious, lastfm and BX have only implicit feedback. In particular, Delicious was from the bookmarking and tagging information\footnote{http://www.delicious.com}, in which each URL was bookmarked by at least 3 users. Lastfm represents music artist listening information\footnote{ http://www.last.fm }, in which each music artist was listened to by at least 10 users and each user listened to at least 5 artists. BX is derived from the Book-Crossing dataset\footnote{http://www.informatik.uni-freiburg.de/~cziegler/BX/} such that only implicit interactions were contained and each book was read by at least 10 users. 

%FilmTrust, Netflix and Yahoo contain multi-value ratings. Specifically, FilmTrust is a dataset crawled from the entire FilmTrust website that allows users to share and assign numerical ratings to movies\footnote{http://www.librec.net/datasets.html}. The Netflix is derived from Netflix Prize dataset\footnote{http://www.netflixprize.com/} and each user rated at least 10 movies. The Yahoo dataset is a subset obtained from Yahoo!Movies user ratings\footnote{http://webscope.sandbox.yahoo.com/catalog.php?datatype=r}. In this dataset, each user rated at least 5 movies and each movie was rated by at least 3 users.
FilmTrust, Netflix and Yahoo contain multi-value ratings. Specifically, FilmTrust is a dataset crawled from the entire FilmTrust website\footnote{http://www.librec.net/datasets.html}. The Netflix is derived from Netflix Prize dataset\footnote{http://www.netflixprize.com/} and each user rated at least 10 movies. The Yahoo dataset is a subset obtained from Yahoo!Movies user ratings\footnote{http://webscope.sandbox.yahoo.com/catalog.php?datatype=r}. In this dataset, each user rated at least 5 movies and each movie was rated by at least 3 users.

\begin{table*}[!ht]
\begin{center}
\begin{threeparttable}[b]
\caption{Comparison of Top-$N$ recommendation algorithms}
\label{tab:comp}
\renewcommand{\arraystretch}{.6}
%\small 
%\begin{large}
%\begin{sc}
%\resizebox{.48\textwidth}{!}{
%\centering
\begin{tabular}{llllllllllllll}
\hline
\multirow{2}{*}{method} &
\multicolumn{6}{c}{Delicious} &
\multicolumn{1}{c}{}&
\multicolumn{6}{c}{lastfm} \\
%\multicolumn{2}{c|}{\multirow{2}{*}{Multi-Row and Col}} \\
\cline{2-7} \cline{9-14} 
  & \multicolumn{4}{c}{params} &\multicolumn{1}{c}{HR}  & \multicolumn{1}{c}{ARHR} &\multicolumn{1}{c}{}&\multicolumn{4}{c}{params} &\multicolumn{1}{c}{HR}  & \multicolumn{1}{c}{ARHR} \\ 
\hline
ItemKNN&300&-&-&-&0.300&0.179&&100&-&-&-&0.125&0.075\\
PureSVD&1000&10&-&-&0.285&0.172&&200&10&-&-&0.134&0.078\\
WRMF&250&5&-&-&0.330&0.198&&100&3&-&-&0.138&0.078\\
BPRKNN&1e-4&0.01&-&-&0.326&0.187&&1e-4&0.01&-&-&0.145&0.083\\
BPRMF&300&0.1&-&-&0.335&0.183&&100&0.1&-&-&0.129&0.073\\
GWNMF&10&5&10&-&0.340&0.211&&20&50&10&-&0.137&0.077\\
SLIM&10&1&-&-&0.343&\bf{0.213}&&5&0.5&-&-&0.141&0.082\\
Our&0.01&0.001&-&-&\bf{0.345}&0.194&&40&10&-&-&\bf{0.198}&\bf{0.085}\\
\hline\hline
\multirow{2}{*}{method} &
\multicolumn{6}{c}{BX} &
\multicolumn{1}{c}{}&
\multicolumn{6}{c}{FilmTrust} \\
%\multicolumn{2}{c|}{\multirow{2}{*}{Multi-Row and Col}} \\
\cline{2-7} \cline{9-14} 
  & \multicolumn{4}{c}{params} &\multicolumn{1}{c}{HR}  & \multicolumn{1}{c}{ARHR} &\multicolumn{1}{c}{}&\multicolumn{4}{c}{params} &\multicolumn{1}{c}{HR}  & \multicolumn{1}{c}{ARHR} \\ 
\hline
ItemKNN&400&-&-&-&0.045&0.026&&5&-&-&-&0.583&0.352\\
PureSVD&3000&10&-&-&0.043&0.023&&20&10&-&-&0.601&0.369\\
WRMF&400&5&-&-&0.047&0.027&&30&2&-&-&0.604&0.371\\
BPRKNN&1e-3&0.01&-&-&0.047&0.028&&4e-3&1e-3&-&-&0.625&0.391\\
BPRMF&400&0.1&-&-&0.048&0.027&&100&0.5&-&-&0.610&0.375\\
GWNMF&10&100&5&-&0.049&0.027&&15&5&2&-&0.615&0.380\\
SLIM&20&0.5&-&-&0.050&0.029&&10&15&-&-&0.628&0.397\\
Our&0.01&0.01&-&-&\bf{0.060}&\bf{0.030}&&1e-4&1e-5&-&-&\bf{0.651}&\bf{0.405}\\
\hline\hline
\multirow{2}{*}{method} &
\multicolumn{6}{c}{Netflix} &
\multicolumn{1}{c}{}&
\multicolumn{6}{c}{Yahoo} \\
\cline{2-7} \cline{9-14} 
  & \multicolumn{4}{c}{params} &\multicolumn{1}{c}{HR}  & \multicolumn{1}{c}{ARHR} &\multicolumn{1}{c}{}&\multicolumn{4}{c}{params} &\multicolumn{1}{c}{HR}  & \multicolumn{1}{c}{ARHR} \\ 
\hline
ItemKNN&200&-&-&-&0.156&0.085&&300&-&-&-&0.318&0.185\\
PureSVD&500&10&-&-&0.158&0.089&&2000&10&-&-&0.210&0.118\\
WRMF&300&5&-&-&0.172&0.095&&100&4&-&-&0.250&0.128\\
BPRKNN&2e-3&0.01&-&-&0.165&0.090&&0.02&1e-3&-&-&0.310&0.182\\
BPRMF&300&0.1&-&-&0.140&0.072&&300&0.1&-&-&0.308&0.180\\
GWNMF&5&30&50&-&0.169&0.080&&20&200&50&-&0.313&0.183\\
SLIM&5&1.0&-&-&\bf{0.173}&\bf{0.098}&&10&1&-&-&0.320&0.187\\
Our&0.02&0.01&-&-&0.166&0.078&&5e-3&5e-6&-&-&\bf{0.379}&\bf{0.194}\\
\hline

\end{tabular}
%\multicolumn{14}{1}{
%\begin{minipage}{6.5cm}
%\resizebox{.48\textwidth}{!}{
 \begin{tablenotes}
      \tiny
%\noindent
      \item[1] The parameters for each method in the table are as follows: ItemKNN: the number of neighbors $k$; PureSVD: the number of singular values and the number of SVD; WRMF: the latent space's dimension and the weight on purchases; BPRKNN: its learning rate and regularization parameter $\lambda$; BPRMF: the latent space's dimension and learning rate; GWNMF: dimension of the latent space, $\lambda$, and $\mu$; SLIM: the $l_2$-norm regularization coefficient $\beta$ and the $l_1$-norm regularization parameter
$\lambda$. Our: item graph regularization parameter $\alpha$ and user graph regularization parameter $\beta$.
 \end{tablenotes}
%\end{minipage}

\end{threeparttable}
%\end{sc}
%\end{large}
\end{center}
\end{table*}

\subsection{Evaluation Methodology}

For fair comparison, we follow the dataset preparation approach used by SLIM \cite{ning2011slim} and adopt the 5-fold cross validation. For each fold, a dataset is split into training and test sets by randomly selecting one non-zero entry for each user and putting it in the test set, while using the rest of the data for training. Then a ranked list of size-$N$ items for each user is produced. We subsequently evaluate the method by comparing the ranked list of recommended items with the item in the test set. In the following results presented in this paper, $N$ is equal to 10 by default.

For Top-$N$ recommendation, the most direct and
meaningful metrics are hit-rate (HR) and the average reciprocal hit-rank (ARHR) \cite{deshpande2004item}, since the users only care if a short recommendation list contains the items of interest or not rather than a very long recommendation list. 
HR is defined as
%\begin{equation}
$HR=\frac{\#\textrm{hits}}{\#\textrm{users}}$,
%\end{equation}
where \#hits is the number of users whose item in the testing set is contained (i.e., hit) in the size-$N$ recommendation list, and \#users is the total number of users. %The maximum HR value is 1.0, which means that the algorithm is able to always recommend the hidden item correctly. An HR value of 0.0 indicates that the algorithm is not able to recommend any of the hidden items. 
%A drawback of HR is that it treats all hits equally. That is HR doesn't consider where they appear in the Top-$N$ list. ARHR resolves this issue by rewarding each hit based on its place in the Top-$N$ list, which
ARHR is defined as:
%\begin{equation}
$ARHR=\frac{1}{\#\textrm{users}}\sum_{i=1}^{\#\textrm{hits}}\frac{1}{p_i}$,
%\end{equation}
where $p_i$ is the position of the $i$-th hit in the ranked Top-$N$ list. In this metric, hits that occur earlier in the ranked list are weighted higher than those occur later, and thus ARHR indicates how strongly an item is recommended. %The highest value of ARHR is equal to HR which occurs when all hits occur in the first position, and the lowest value is equal to HR/$N$ when all hits occur in the last position of the list.

\begin{figure*}[!ht]
\begin{center}
%\resizebox{\textwidth}{!}{
\subfigure[Delicious]{\includegraphics[width=.3\textwidth]{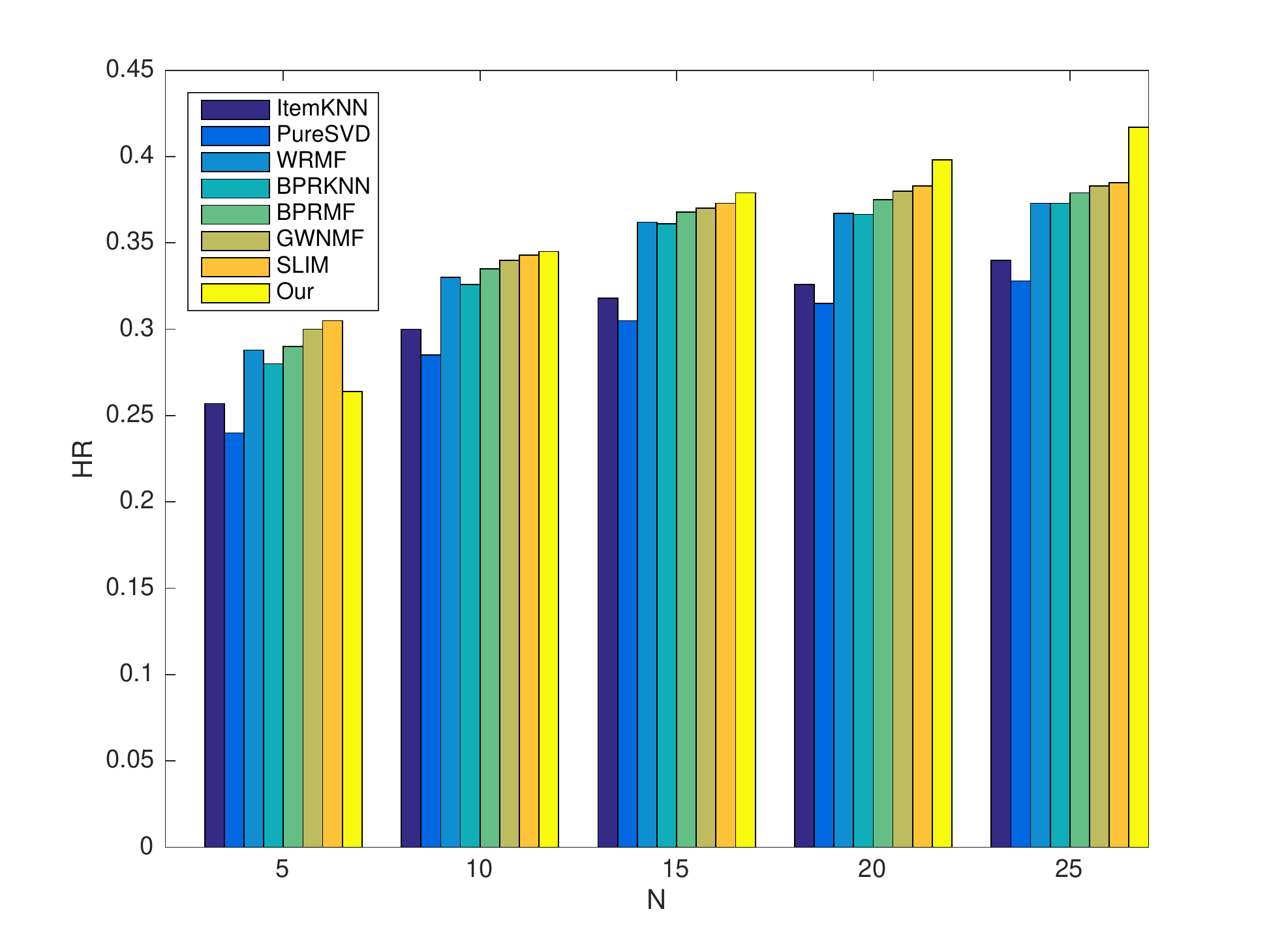}}
\subfigure[lastfm]{\includegraphics[width=.3\textwidth]{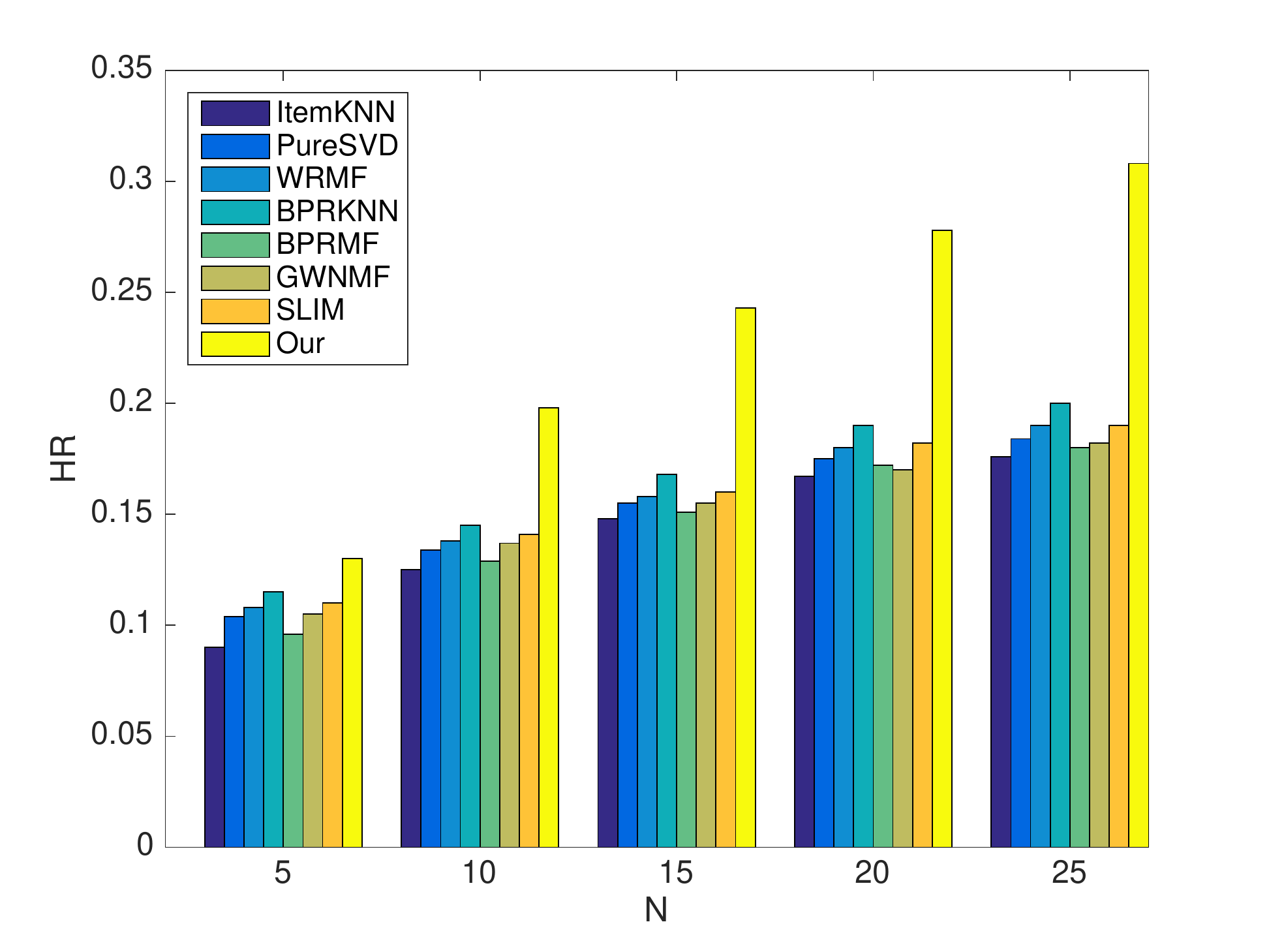}}
\subfigure[BX]{\includegraphics[width=.3\textwidth]{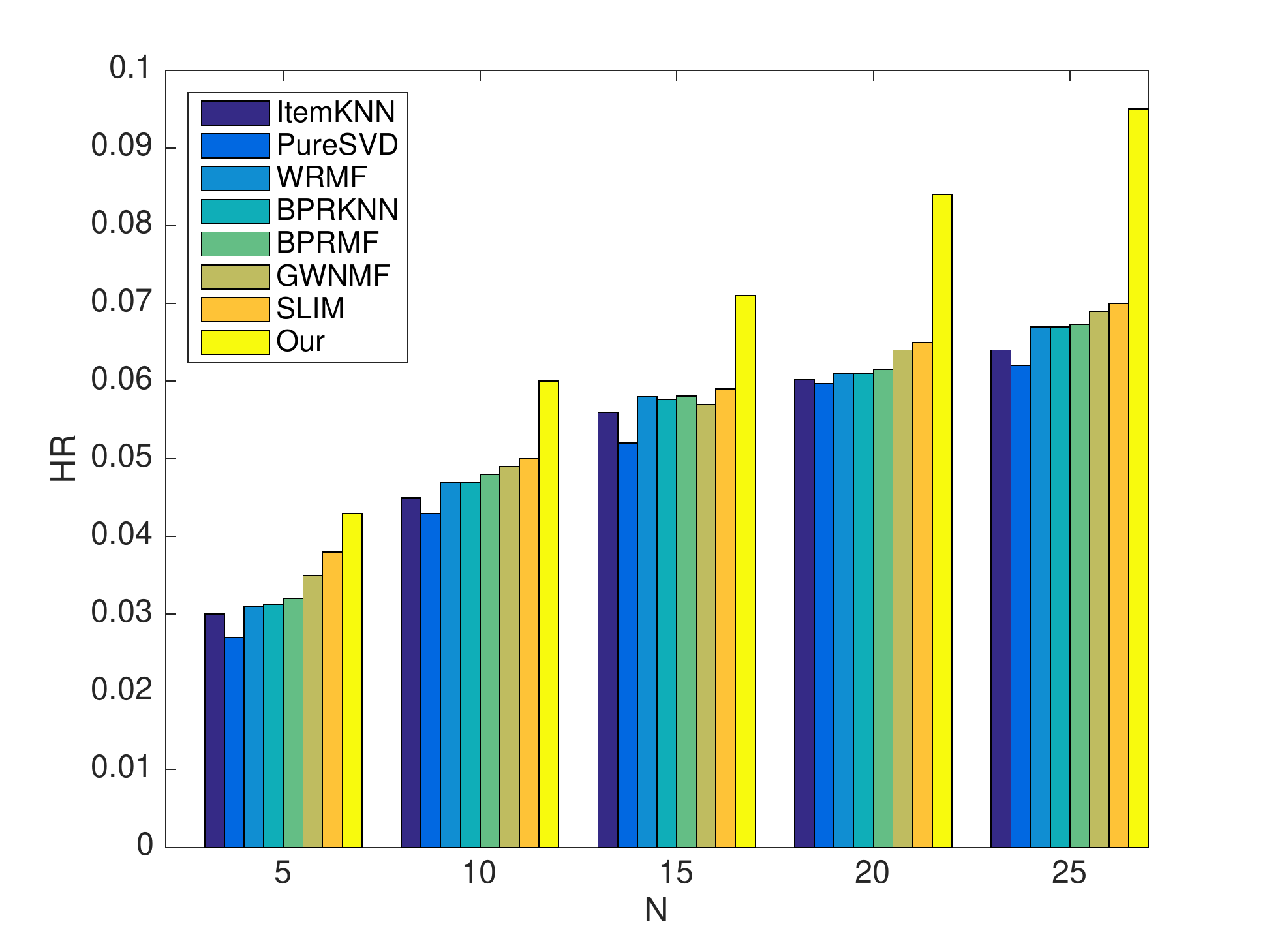}}\\
\subfigure[FilmTrust]{\includegraphics[width=.3\textwidth]{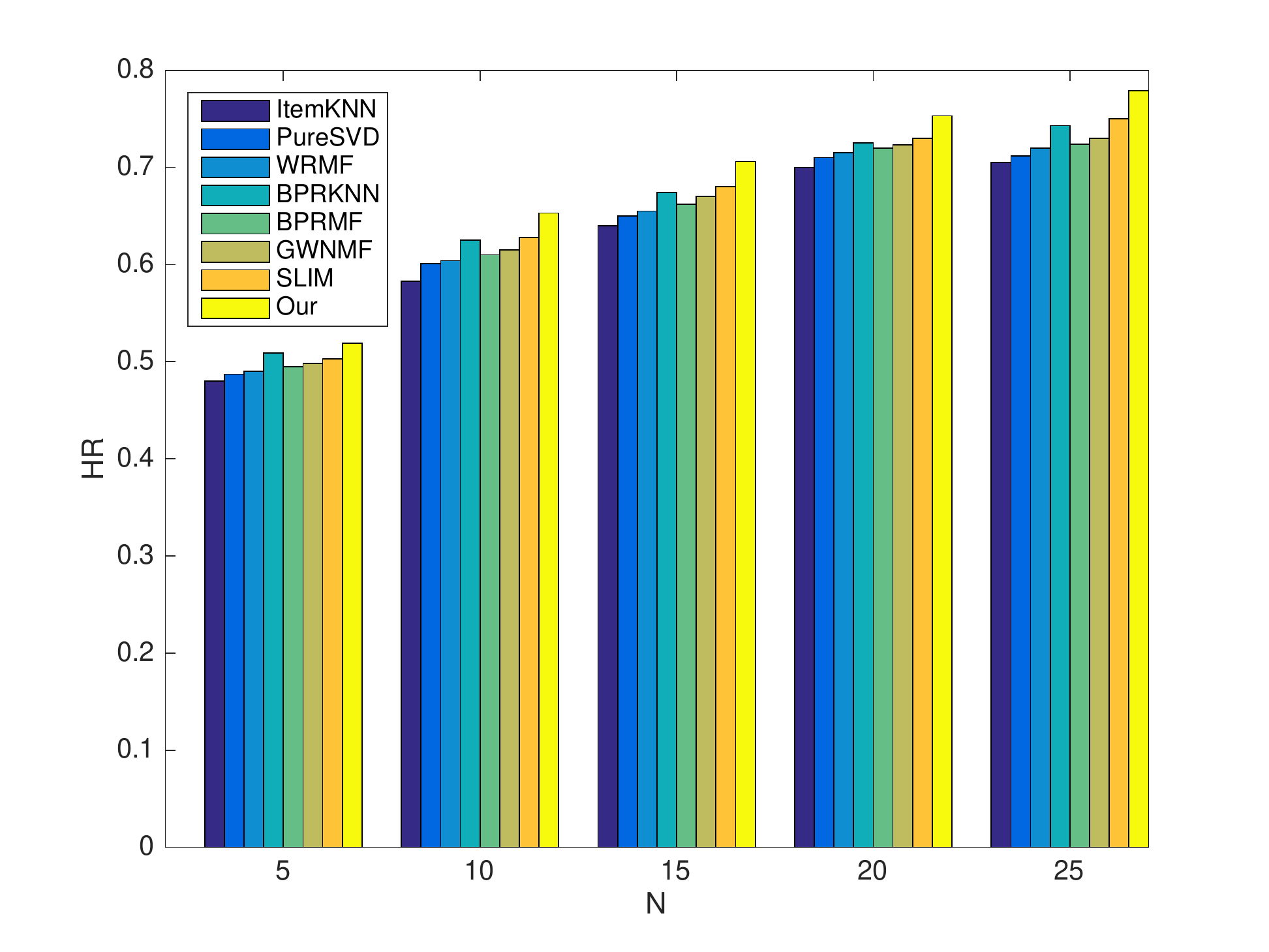}}
\subfigure[Netflix]{\includegraphics[width=.3\textwidth]{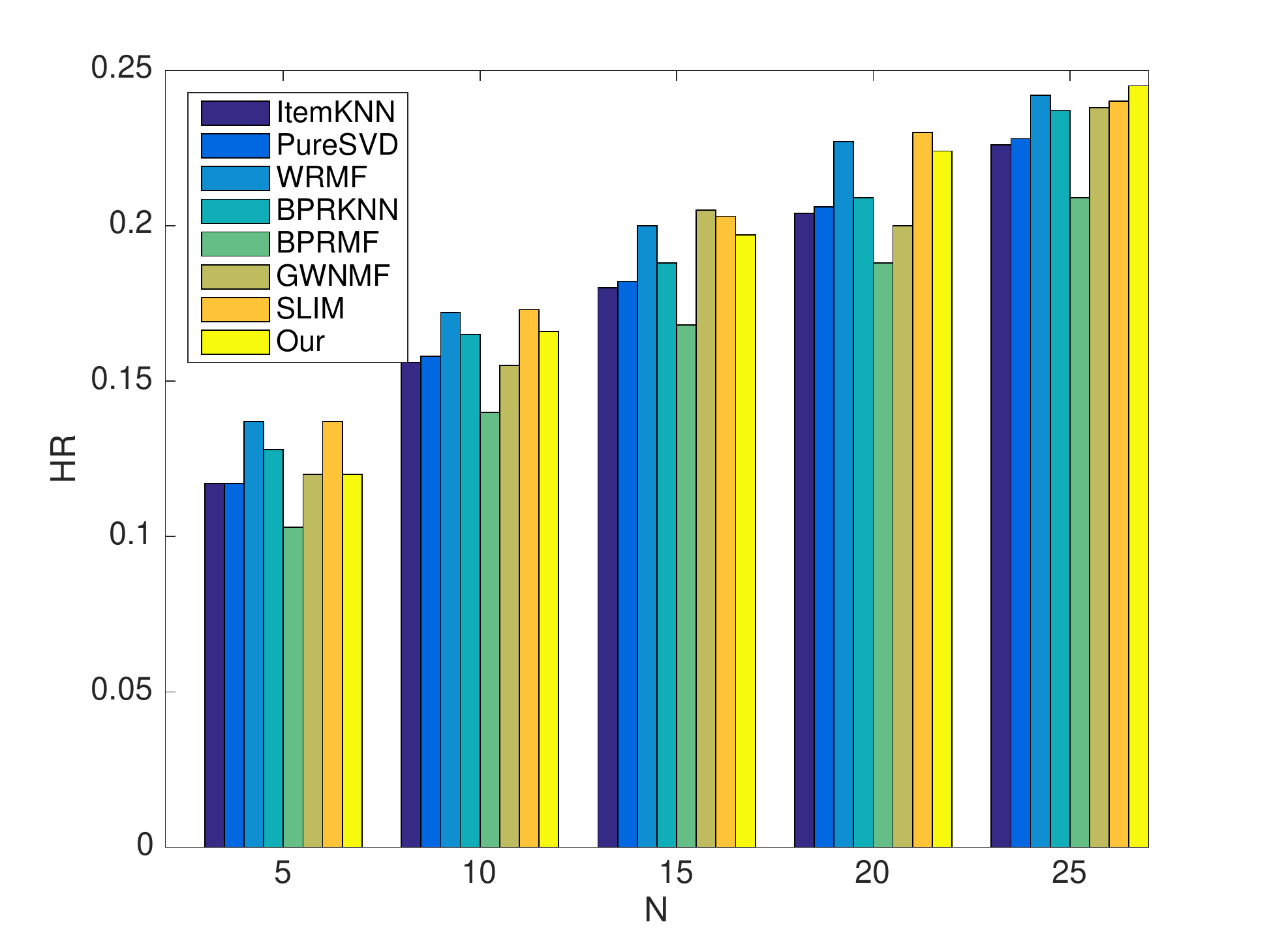}}
\subfigure[Yahoo]{\includegraphics[width=.3\textwidth]{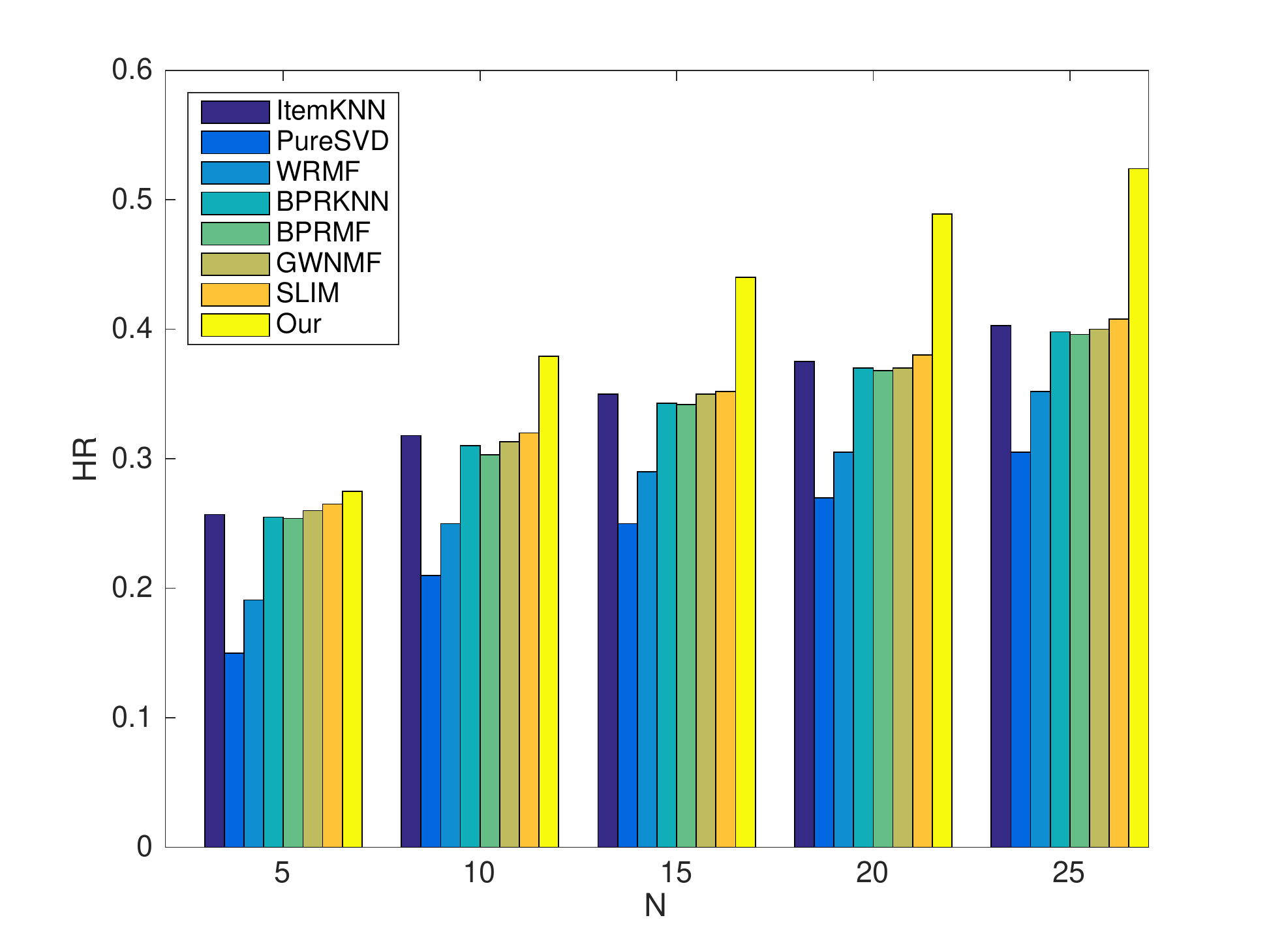}}
\caption{ Performance versus Different Values of $N$.}
\label{fig:differentN}
%\vspace{-.4cm}
\end{center}
\end{figure*}
%\subsection{Comparison Algorithms}
%To extensively assess the performance of our proposed approach, we compare with seven state-of-the-art Top-$N$ recommendation algorithms. These methods include the item neighborhood-based collaborative filtering method ItemKNN \cite{deshpande2004item}; three MF-based methods PureSVD \cite{cremonesi2010performance}, WRMF \cite{hu2008collaborative}, and GWNMF \cite{gu2010collaborative}; SLIM \cite{ning2011slim}; two ranking/retrieval criteria based methods BPRMF and BPRKNN \cite{rendle2009bpr}. These algorithms form a good set of methods to compare with and evaluate the proposed approach.% In addition, their implementations are eithor available from existing library, e.g., LibRec\footnote{http://www.librec.net/index.html}, or graciously provided by their authors. %The experiments are conducted using the platform Intel Xeon E3-1240 3.40GHz CPU with 8GB RAM. We implement our algorithm in MATLAB (R2012a) and   complete code for our algorithm will be available at ************. 
\begin{figure*}[h]
\centering
\subfigure[FilmTrust]{\includegraphics[width=.245\textwidth]{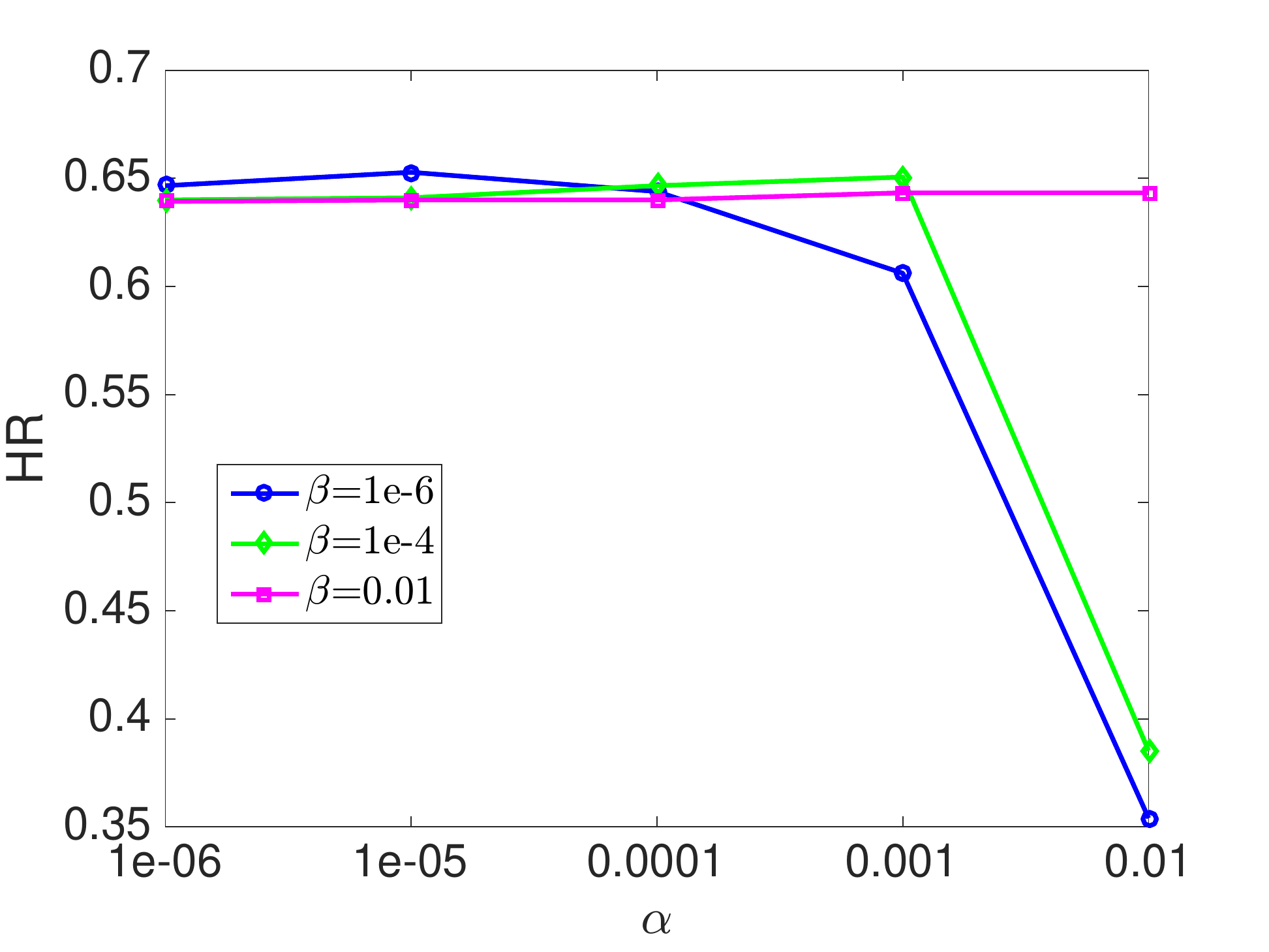}}
\subfigure[FilmTrust]{\includegraphics[width=.245\textwidth]{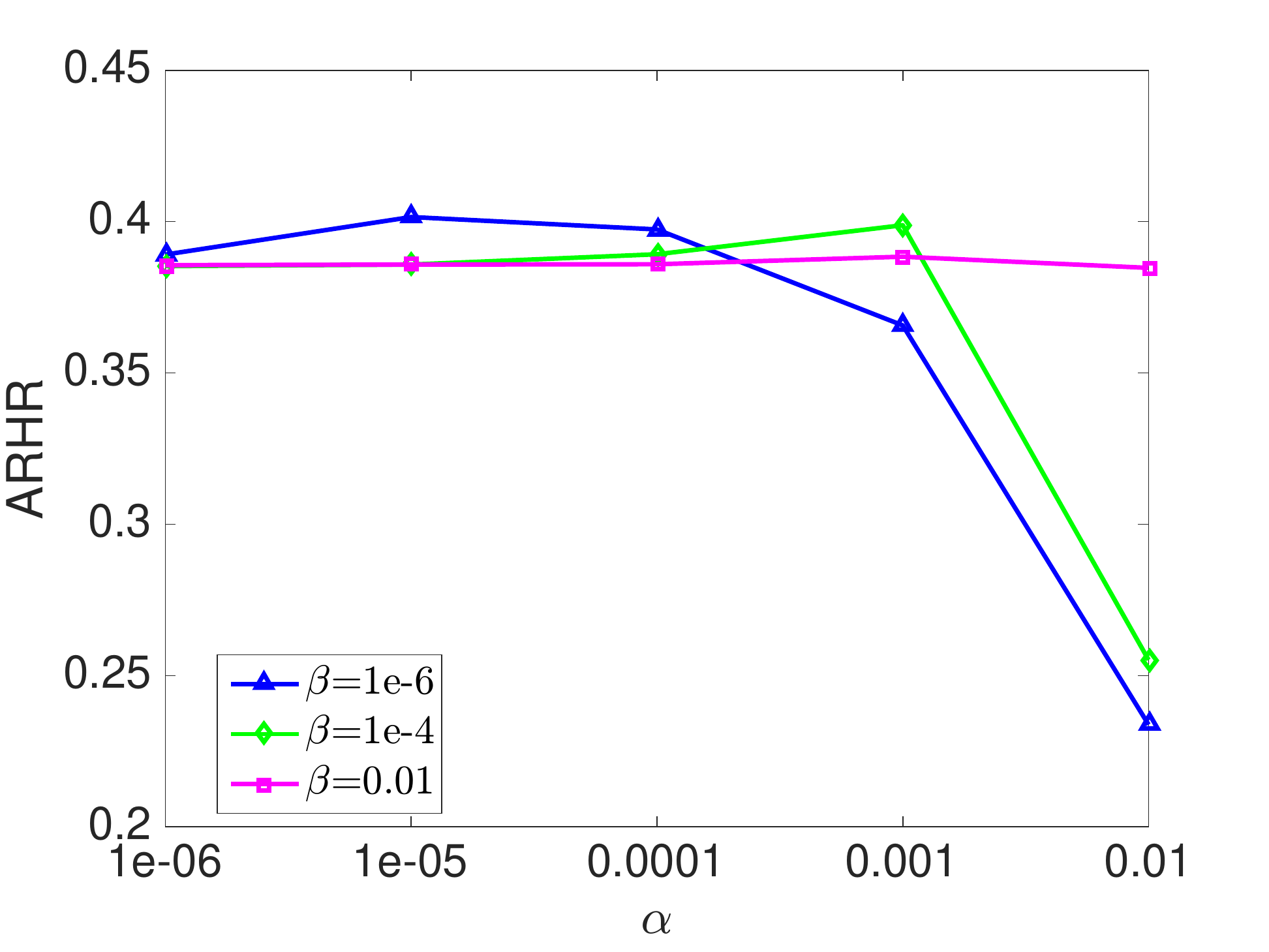}}
\subfigure[Yahoo]{\includegraphics[width=.245\textwidth]{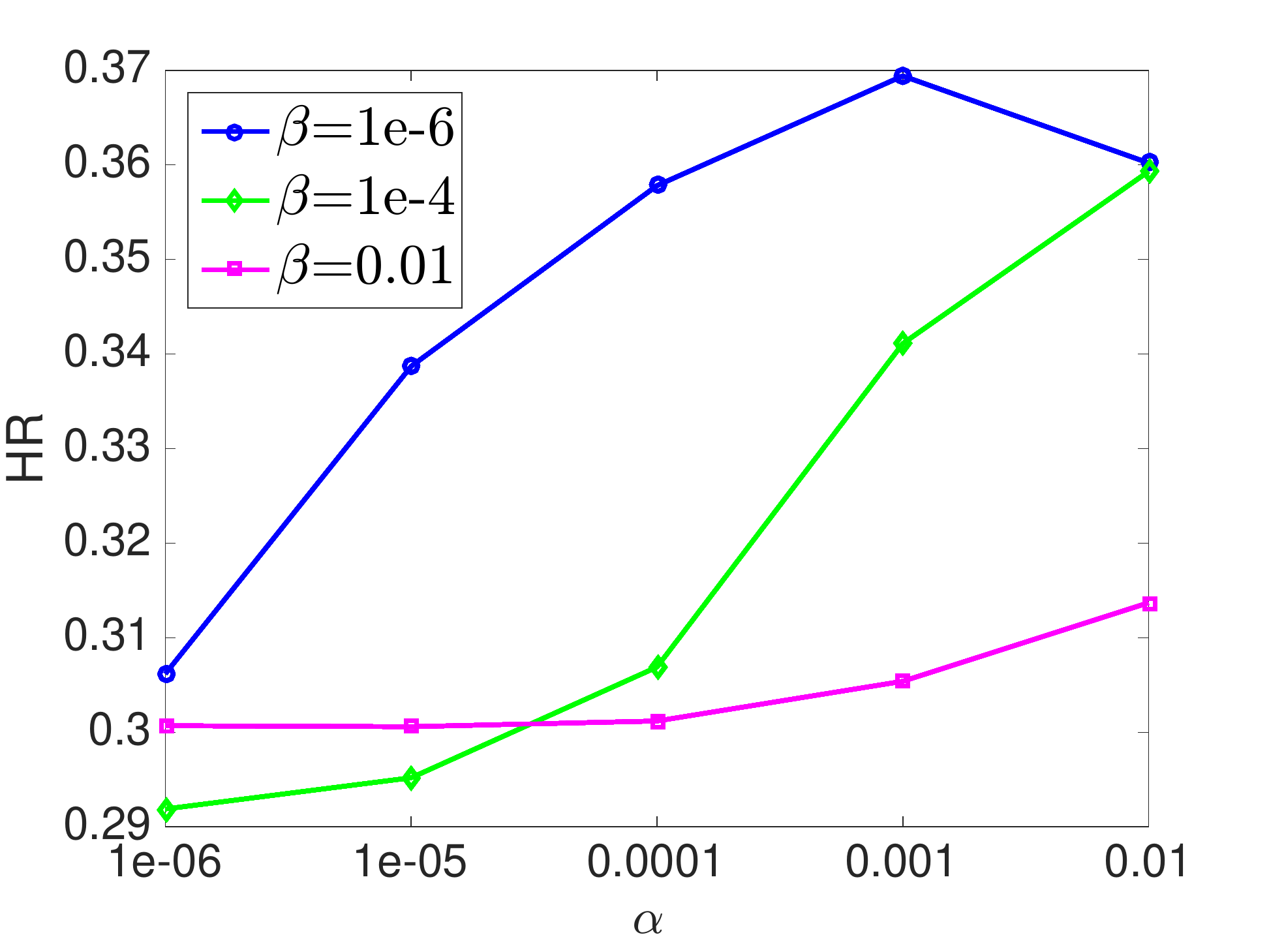}}
\subfigure[Yahoo]{\includegraphics[width=.245\textwidth]{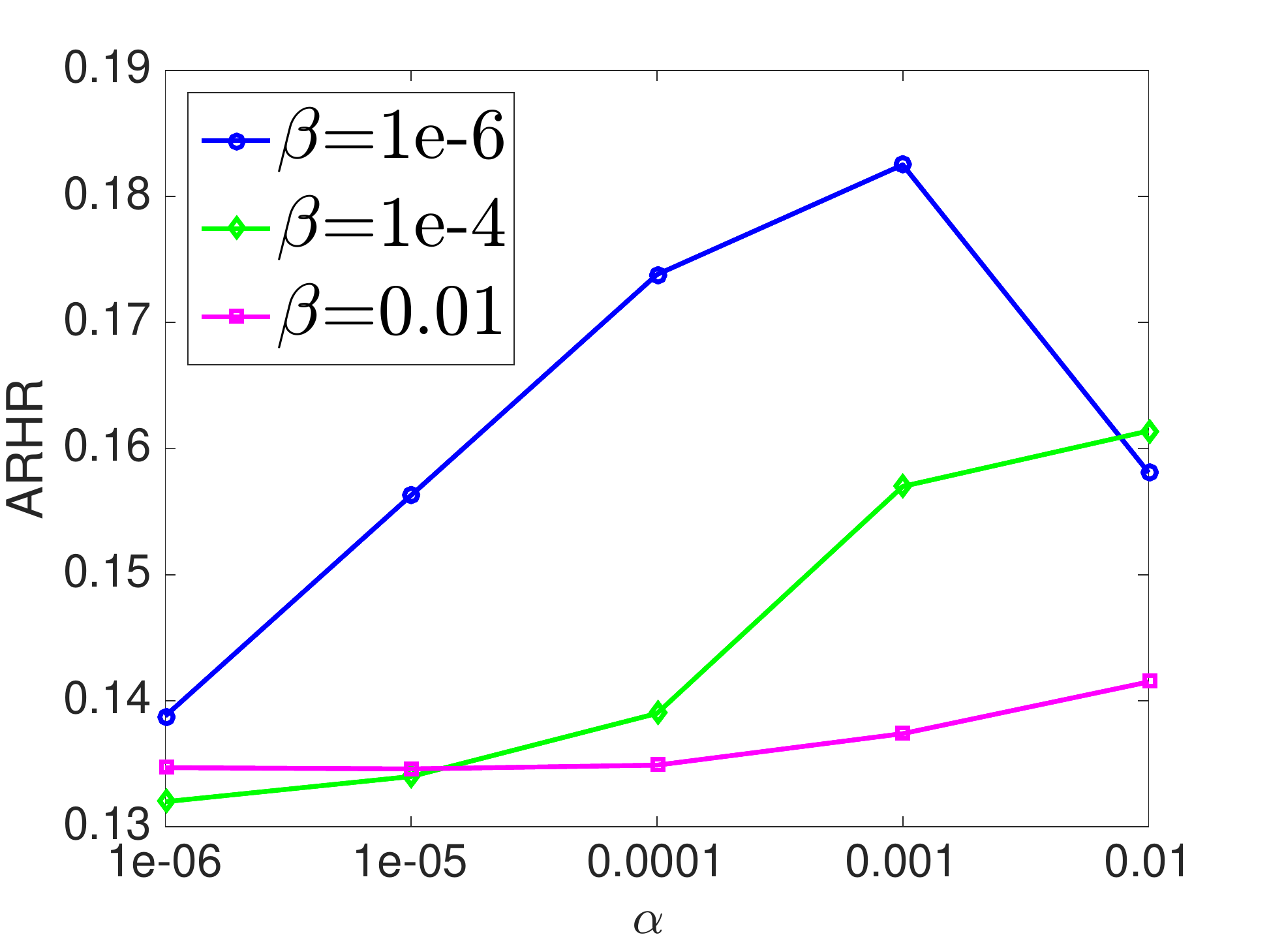}}
%\subfigure[ARHR with varying $\alpha$ and $\beta$]{\includegraphics[width=.4\textwidth]{arhr.pdf}}
\caption{ Influence of $\alpha$ and $\beta$ on HR and ARHR.}
\label{parameter}
\end{figure*}

 %\vfill\eject
\section{Experimental Results}
\label{discuss}

%With the setup in the previous section, this section demonstrates that the algorithm is capable of effectively recommending relevant items to the users. 
\subsection{Top-N Recommendation Performance}
We use 5-fold cross-validation to choose parameters for all competing methods and report their best performance in Table \ref{tab:comp}. It can be seen that the HR improvements achieved by our method against the next best performing scheme (i.e., SLIM) are quite substantial on lastfm, Yahoo, BX, FilmTrust datasets\footnote{Code is available at https://github.com/sckangz/CIKM16}. For Delicious and Netflix datasets, our performance is close to the best performance of other methods. In most cases, there is no much difference among other state-of-the-art methods in terms of HR.
Figure \ref{fig:differentN} shows the performance in HR of various methods for different values of $N$(i.e., 5, 10, 15, 20 and 25) on all six datasets. Our method works the best in most cases.

\subsection{Parameter Effects}
Our model involves two trade-off parameters $\alpha$ and $\beta$, which dictate how strongly item and user neighborhoods and structure information contribute to the objective and performance.
In Figure \ref{parameter}, we depict the effects of different $\alpha$ and $\beta$ values on HR and ARHR for dataset FilmTrust and Yahoo. The search for $\alpha$ ranges from 1e-6 to 1e-2 with points from $\{\text{1e-6, 1e-5, 1e-4, 1e-3, 1e-2}\}$, the search points for $\beta$ are from $\{\text{1e-6, 1e-4, 1e-2}\}$. As can be seen from all figures, our algorithm performs well over a wide range of $\alpha$ and $\beta$ values. HR and ARHR share the same trend with varying $\alpha$ and $\beta$. Specifically, when $\alpha$ is small, HR and ARHR both increase with $\alpha$. After a certain point, they begin to decrease. For FilmTrust, the performance with $\beta=0.01$ is very stable with respect to $\alpha$. This suggests that user-user similarity dominates the FilmTrust dataset.

\subsection{Matrix Reconstruction}
To show how our method reconstructs the user-item matrix, we compare it with the method of next best performance, SLIM, on FilmTrust. The density of FilmTrust is 1.14\% and the mean for those non-zero elements is 2.998. The reconstructed matrix $\hat{X}_{SLIM}$ from SLIM has a density of 83.21\%. For those 1.14\% non-zero entries in $X$, $\hat{X}_{SLIM}$ recovers 99.69\% of them and their mean value is 1.686. In contrast, the reconstructed matrix by our proposed algorithm has a density of 91.7\%. For those 1.14\% non-zero entries in $X$, our method recovers all of them with a mean of 2.975. These facts suggest that our method better recovers $X$ than SLIM. In other words, SLIM loses too much information. This appears to explain the superior performance of our method.  

In fact, above analysis is equivalent to the two widely used prediction accuracy metrics: Mean Absolute Error (MAE) and Root Mean Squared Error (RMSE). Since our method can recover the original ratings much better than SLIM, our algorithm gives lower MAE and RMSE. This conclusion is consistent with our HR and ARHR evaluation.
\begin{table}[!htbp]
\centering
\caption{Results with Different Similarity Metrics}
\label{simcomp}
\renewcommand{\arraystretch}{.6}
\begin{tabular}{|l|c|c|c|c|}%lclclclcl}
\hline
\multirow{2}{*}{Datasets} &
\multicolumn{2}{c|}{HR} &
\multicolumn{2}{c|}{ARHR} \\
%\multicolumn{2}{c|}{\multirow{2}{*}{Multi-Row and Col}} \\
\cline{2-3} \cline{4-5} 
  & \multicolumn{1}{c|}{Jaccard} &\multicolumn{1}{c|}{Cosine}  & \multicolumn{1}{c|}{Jaccard} &\multicolumn{1}{c|}{Cosine}\\%&\multicolumn{4}{c}{params} &\multicolumn{1}{c}{HR}  & \multicolumn{1}{c}{ARHR} \\ 
\hline
Delicious&\bf{0.345}&0.330&0.194&\bf{0.202}\\
\hline
lastfm&0.198&\bf{0.204}&0.085&\bf{0.087}\\
\hline
BX&\bf{0.060}&0.057&\bf{0.030}&0.024\\
\hline
\end{tabular}

\end{table}
 \vfill\eject
\subsection{Graph Construction}
\begin{figure}[h]
\centering
\includegraphics[width=.22\textwidth]{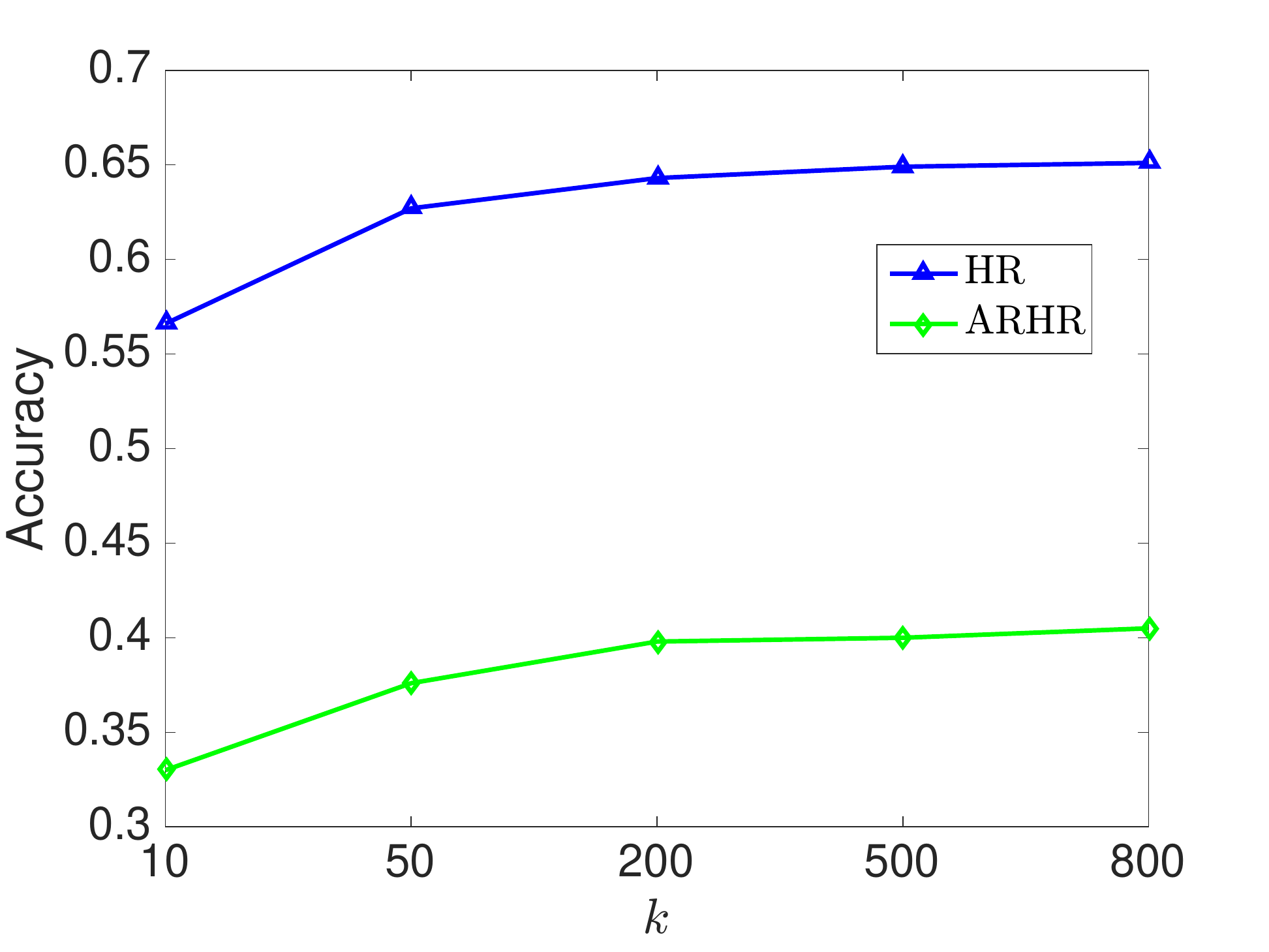}
\caption{ Influence of neighborhood size $k$ on recommendation accuracy for FilmTrust dataset.}
\label{neibor}
\end{figure}
As we discussed previously, similarity is an important ingredient of graph construction. In many recommendation algorithms, the similarity computation is crucial to the recommendation quality \cite{guo2013novel}. 
To demonstrate the importance of similarity metric, we use the cosine measure rather than the Jaccard coefficient to measure the similarity in binary datasets. We compare the results in Table \ref{simcomp}. As demonstrated, for lastfm dataset, HR and ARHR increase after we adopt the cosine similarity. However, for Delicious and BX dataset, the Jaccard coefficient works better. Therefore, the difference of final results can be big for certain datasets with different similarity measures. We expect that the experimental results in Table \ref{tab:comp} can be further enhanced if one performs a more careful analysis of $L_r$ and $L_c$. For example,
it has been reported that normalizing the similarity scores can improve the performance \cite{deshpande2004item}. Also, a number of new similarity metrics have recently been proposed, e.g., \cite{liu2014new}, which may be also exploited.

\begin{table}%[!htbp]
%\centering
\caption{Results by using different graphs}
\label{singlegraph}
\renewcommand{\arraystretch}{.6}{
\begin{tabular}{|c|c|c|c|}%lclclclcl}
\hline
%%\multirow{1}{}{HR} &
%%\multicolumn{1}{c|}{User Graph} &
%%\multicolumn{1}{c|}{Item Graph} &
%%\multicolumn{1}{c|}{User-Item Graph} \\
%%\multicolumn{2}{c|}{\multirow{2}{*}{Multi-Row and Col}} \\
%\cline{1-4}
%  %& \multicolumn{1}{c|}{Jaccard} &\multicolumn{1}{c|}{Cosine}  & \multicolumn{1}{c|}{Jaccard} &\multicolumn{1}{c|}{Cosine}\\%&\multicolumn{4}{c}{params} &\multicolumn{1}{c}{HR}  & \multicolumn{1}{c}{ARHR} \\ 
&User Graph&Item Graph&User-Item Graph\\
\hline
FilmTrust&0.638&0.625&0.651\\
\hline
Yahoo&0.303&0.379&0.379\\
\hline
\end{tabular}}
\end{table}
Another important parameter is the neighborhood size $k$, which cannot be known a priori \cite{huang2015new}. For some small datasets, setting a small
$k$ may not include all useful neighbors and would infer incomplete relationships. In practice, a large number of ratings from similar users or similar items are not available, due to the sparsity inherent to rating data. We just use the fully connected graph in our experiments. To demonstrate this, we test the effects of neighborhood size $k$ with values $\{10, 50, 200, 500, 800\}$ on FilmTrust data. As can be seen from Figure \ref{neibor}, the neighborhood size indeed influences the performance of our proposed recommendation method. Specifically, the performance keeps increasing as $k$ increases when $k$ is small compared to the size of dataset, then the performance keeps almost the same as the final accuracy obtained in Table \ref{tab:comp} as $k$ it becomes larger. This conforms that a small neighborhood size can not capture all similarity information. %On the other hand, a dense graph Laplacian matrix may increase data storages and computations. Therefore, we can seek a balance between recommendation quality and the number of  %From another point of view, the neighborhood size has some influence in the computational time.%, as displayed in Figure \ref{time}. As we said before, the time will increase with the dense of graph Laplacian. Therefore, to speed up the algorithm in reality, we can make $L_r$ and $L_c$ sparse by just keeping similarities between each user/item and its small $k$ most similar users/items. 
\subsection{Effects of User and Item Graphs}
While the overall improvements are impressive, it would be interesting to see more fine-grained analysis of the impact of user-user and item-item similarity graphs. We use FilmTrust and Yahoo datasets as examples to show the effects of user and item graphs. Table \ref{singlegraph} summarizes the HR values obtained with user graph, item graph, and both user and item graph. It demonstrates that we are able to obtain the best performance when we combine user and item graph. Thus neighborhood information of users and items can alleviate the problem of data sparsity by taking advantage of structural information more extensively, which in turn benefits the recommendation accuracy. %In the popular matrix factorization based methods, the factorization of the user-item matrix may has many possible solutions due to the highly sparse rating matrix \cite{gu2010collaborative}. %The user and item similarity information is complementary to each.

%It is important to point out that our method is much faster than LorSLIM. Among the six datasets, FilmTrust and lastfm datasets have the smallest and largest sizes $943\times 1682$, $8813\times 6038$, respectively. Our method needs 5.7s and 1385s, respectively, on these two datasets, while LorSLIM takes 617s and 32974s. %The time is measured on the same machine with an Intel Xeon E3-1240 3.40GHz CPU that has 4 cores and 8GB memory, running Ubuntu and Matlab (R2014a). We believe that our speed can be further improved by many other packages.
\section{Conclusion}
\label{conclude}
In this paper, we address the demands for high-quality recommendation on both implicit and explicit feedback datasets. We reconstruct the user-item matrix by fully exploiting the similarity information between users and items concurrently. Moreover, the reconstructed data matrix also respects the manifold structure of the user-item matrix. We conduct a comprehensive set of experiments and compare our method with other state-of-the-art Top-$N$ recommendation algorithms. The results demonstrate that the proposed algorithm works effectively. %improve prediction accuracy considerably. %In addition, our proposed method involves solving a Sylvester equation only, which makes it efficient for large-scale datasets. 
Due to the simplicity of our model, there is much room to improve. For instance, our model can be easily extended to include side information (e.g., user demographic information, item's genre, social trust network) by utilizing  the graph representation. In some cases, external information is more informative than the neighborhood information. %In the future, we will explore further along this line.
\section{Acknowledgments}
This work is supported by the U.S. National Science Foundation under Grant IIS 1218712, National Natural Science Foundation of China under grant 11241005, and Shanxi Scholarship Council of China 2015-093. Q. Cheng is the corresponding author.
\bibliographystyle{abbrv}
\bibliography{recom}  % sigproc.bib is the name of the Bibliography in this case

\end{document}